\pdfoutput=1

\documentclass{article}
\usepackage[utf8]{inputenc}
\usepackage[a4paper, margin=25mm]{geometry}

\usepackage[fontsize=10.5pt]{fontsize}
\usepackage{mathptmx}
\usepackage{helvet}

\usepackage{amsfonts,amsmath,amsthm,amssymb}
\usepackage{bm}
\usepackage{comment}
\usepackage[sort&compress, numbers, round]{natbib}

\usepackage{lipsum}

\usepackage[colorlinks=true, urlcolor=blue]{hyperref}
\usepackage{cleveref}
\usepackage{afterpage}
\usepackage{graphicx}

\usepackage{placeins}

\usepackage{xcolor}
\definecolor{headertextcolor}{RGB}{89,89,89}
\definecolor{loaclr}{RGB}{152, 78, 163}
\definecolor{mclr}{RGB}{255, 127, 0}

\usepackage{authblk}

\usepackage{manyfoot}
\DeclareNewFootnote{B}[arabic]

\usepackage{indentfirst}

\usepackage[explicit, largestsep]{titlesec}

\titleformat{\section}[block]
    {\fontsize{12}{14.4}\selectfont\sffamily\bfseries}
    {\makebox[1cm][l]{\thesection.}\MakeUppercase{#1}}{0pt}{}
\titleformat{name=\section, numberless}[block]
    {\fontsize{12}{14.4}\selectfont\sffamily\bfseries}
    {\MakeUppercase{#1}}{0pt}{}
\titlespacing*{\section}{0pt}{12pt}{1pt}

\titleformat{\subsection}[block]
  {\selectfont\sffamily\bfseries}
  {\makebox[1cm][l]{\thesubsection}#1}{0pt}{}
\titlespacing*{\subsection}{0pt}{6pt}{\parskip}
 
\titleformat{\subsubsection}[block]
  {\selectfont\sffamily\bfseries}
  {\makebox[1cm][l]{\thesubsubsection}#1}{0pt}{}
\titlespacing*{\subsubsection}{0pt}{3pt}{\parskip}

\usepackage{fancyhdr}
\fancypagestyle{firststyle}{
    \fancyhf{}
    \fancyhead[C]{%
        \begin{minipage}{1.85in}%
            \includegraphics[width=1.63in]{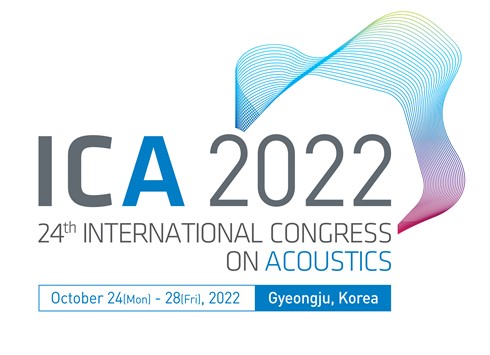}\\
        \end{minipage}
        \begin{minipage}[c]{3.34in}%
            \fontsize{10}{12}\selectfont\sffamily\bfseries
            \color{headertextcolor}
            PROCEEDINGS of the\\
            24th International Congress on Acoustics\\\\
            October 24 to 28, 2022 in Gyeongju, Korea
        \end{minipage}
    }

    \fancyfoot[C]{%
        \includegraphics[width=1.1in]{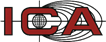}
        \hskip0.5in
        \includegraphics[width=0.53in]{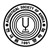}
    }

}

\makeatletter
\renewcommand{\maketitle}{
    \vspace*{.75in}
    \begin{center}
        \fontsize{14}{16.8}\selectfont\sffamily\bfseries
        \@title
    \end{center}
    \begin{center}
        \@author
    \end{center}
    \enlargethispage{-0.25in}
}
\makeatother

\makeatletter
\renewcommand{\abstractname}{{\sffamily\bfseries ABSTRACT\\}}
\renewcommand{\abstract}{\noindent\abstractname}
\g@addto@macro\endabstract{\vspace{1em}\par}
\makeatother

\newcommand{\keywords}[1]{\noindent Keywords: #1}

\makeatletter
\def\@makecaption#1#2{%
  \vskip\abovecaptionskip
  \hb@xt@\hsize{\hfil#1 -- #2\hfil}%
  \vskip\belowcaptionskip}
\makeatother

\usepackage{tabularx}
\usepackage{multirow}
\usepackage{booktabs}

\usepackage{siunitx}
\DeclareSIUnit \decibelA {dB(A)}
\DeclareSIUnit \decibelC {dB(C)}
\DeclareSIUnit \soneGF {soneGF}
\DeclareSIUnit \acum {acum}
\DeclareSIUnit \asper {asper}
\DeclareSIUnit \vacil {vacil}
\DeclareSIUnit \tuhms {tuHMS}

\usepackage[style=base]{caption}

\begin{document}

\title{Assessment of a cost-effective headphone calibration procedure for soundscape evaluations}

\author{Bhan LAM}
\author{Kenneth OOI}
\author{Zhen-Ting ONG}
\author{\authorcr Karn N. WATCHARASUPAT}
\author{Trevor WONG}
\author{and Woon-Seng GAN}
\affil{School of Electrical and Electronic Engineering, Nanyang Technological University, Singapore\textsuperscript{1}}
\FootnotetextB{1}{\{bhanlam, wooi002, ztong, ylau01, karn001, trevor.wong, ewsgan\}@ntu.edu.sg}



\maketitle

\begin{abstract}
    To increase the availability and adoption of the soundscape standard, a low-cost calibration procedure for reproduction of audio stimuli over headphones was proposed as part of the global ``Soundscape Attributes Translation Project'' (SATP) for validating ISO/TS~12913-2:2018 perceived affective quality (PAQ) attribute translations. A previous preliminary study revealed significant deviations from the intended equivalent continuous A-weighted sound pressure levels ($L_{\text{A,eq}}$) using the open-circuit voltage (OCV) calibration procedure. For a more holistic human-centric perspective, the OCV method is further investigated here in terms of psychoacoustic parameters, including relevant exceedance levels to account for temporal effects on the same 27 stimuli from the SATP. Moreover, a within-subjects experiment with 36 participants was conducted to examine the effects of OCV calibration on the PAQ attributes in ISO/TS~12913-2:2018. Bland-Altman analysis of the objective indicators revealed large biases in the OCV method across all weighted sound level and loudness indicators; and roughness indicators at \SI{5}{\%} and \SI{10}{\%} exceedance levels. Significant perceptual differences due to the OCV method were observed in about \SI{20}{\%} of the stimuli, which did not correspond clearly with the biased acoustic indicators. A cautioned interpretation of the objective and perceptual differences due to small and unpaired samples nevertheless provide grounds for further investigation.
\end{abstract}

\afterpage{\aftergroup\restoregeometry}

\section{Introduction}

Accurate calibration of recordings and stimuli to desired \citep{Galbrun2013} or in-situ \citep{Aletta2018b} sound pressure levels (SPL) is crucial to laboratory-based soundscape studies, because differences in SPL may influence ratings of perceptual indicators of the soundscapes under study. For example, preference scores of the same road traffic noise recording were found to be lower when it was presented at A-weighted equivalent SPL ($L_{\text{A,eq}}$) of \SI{75}{\decibel}, as compared to when it was presented at \SI{55}{\decibel} \citep{Jeon2011b}. Similarly, significant differences in subjectively-rated perceived loudness and overall soundscape quality were observed when the same hydraulic breaker sound was presented at $L_{\text{A,eq}}$ values of \SI{55}{}, \SI{65}{}, and \SI{75}{\decibel} \citep{Hong2020EffectsQualityKenneth}.

While much effort has been put into standardizing the definitions \citep{InternationalOrganizationforStandardization2014ISOFramework}, measurement \citep{InternationalOrganizationforStandardization2018x}, and analysis \citep{InternationalOrganizationforStandardization2019} of soundscapes via the ISO 12913 series of standards, there appears to be scant detail on exact calibration procedures that should be undertaken to control the SPLs at which stimuli are presented in soundscape studies. Hence, some studies have used ``subjective'' calibration methods, where participants adjusted presentation SPLs on fixed perceptual criteria, whereas others have used ``objective'' calibration methods, where physical equipment is used for the calibration procedure instead of human participants.

Sudarsono et al.\ proposed a ``subjective'' calibration method in \citep{Sudarsono2016Kenneth}, where participants adjusted soundscape recordings to levels that they \textit{thought} matched the in-situ recording conditions (even though they had not experienced those conditions themselves). Most participants adjusted them to levels between \SI{5}{\decibel} to \SI{15}{\decibel} lower than the in-situ levels. 
However, results from a semantic differential analysis of the adjusted recordings against in-situ conditions showed no significant difference between the laboratory and in-situ conditions, which suggests that results could be ecologically valid even if reproduced levels differ from in-situ or ``desired'' levels. In addition, a study by Kothinti et al.\ \citep{Kothinti2021AuditoryStudyKenneth} revealed no significant differences in behavioral salience (the duration which a participant was more attentive to one of two simultaneously-presented stimulli) measurements between crowdsourced data obtained from participants who adjusted stimuli to a comfortable listening level and laboratory data obtained from participants who were presented the same stimuli at levels calibrated by the researchers. 

On the other hand, ``objective'' calibration methods can be automated and may be less tedious to perform since they do not require the involvement of human participants. Open-circuit voltage (OCV) methods, where the same gain adjustment is made to all stimuli using a \textit{separate} reference track based on the root-mean-squared (RMS) voltage across a playback device, are generally quick and economical, since they do not require specialized equipment beyond the playback equipment already needed for stimuli presentation in the first place. An OCV method was first used by Park et al.\ to calibrate tracks prior to the extraction of acoustic parameters for analysis \citep{Park2013Kenneth}, and later independently proposed by Aletta et al.\ for in use presenting a standard set of 27 stimuli for validation of candidate translations of perceptual attributes in the Soundscape Attributes Translation Project (SATP) \citep{Aletta2020Kenneth}, and Park et al.\ also previously used an OCV method was to calibrate tracks prior to the extraction of acoustic parameters for analysis \citep{Park2013Kenneth}. 

Alternatively, methods using head-and-torso simulators (HATS) to account for binaural presentation of stimuli and pinna cues have previously been used by De Coensel et al.\ \citep{Coensel2011Kenneth}, and have also been proposed in \citep{Ooi2021AutomationHeadKenneth} for automated calibration of soundtracks reproduced over headphones or loudspeakers with binaural systems compliant with ISO/TS 12913-2:2018. A GRAS 43AG Ear and Cheek simulator was also used for calibration of stimuli in a study on music perception by McPherson et al.\ \citep{McPherson2020PerceptualIntervalsKenneth}. Although such methods can account for variability in headphone quality and playback devices, the need for specialized equipment may make them cost-prohibitive for some research groups.

Therefore, we aim to quantify the differences (if any) in OCV and HATS calibration methods for stimuli in the context of soundscape studies, and seek to investigate if an OCV method could potentially be a cost-effective alternative to a HATS calibration method. In terms of the objectively-measured SPL, a preliminary study has found differences in SPL between OCV and HATS calibration methods ranging from \SI{2.5}{\decibel} to \SI{8.75}{\decibel}, depending on the stimulus \citep{Lam2022ax}. For this study, we provide an overview of the preliminary study, and extend it with additional objective analysis, as well as subjective analysis utilizing perceptual characteristics according to Part 2 of the Method A questionnaire in ISO/TS 12913-2:2018 \citep{InternationalOrganizationforStandardization2018}, in the context of candidate translations for the Bahasa Melayu version of the questionnaire.

\section{Methodology}

\subsection{Stimuli and calibration}
Both the subjective and objective analysis are based on a set of 27 audio stimuli that were used in SATP. All 27 stimuli were 30-s binaural excerpts recorded with a wearable binaural microphone and a data acqusition device (BHS~II; SQobold, HEAD acoustics GmbH, Herzogenrath, Germany) according to the protocol described in \citep{Mitchell2020}. These 27 excerpts were then calibrated for headphone playback using one of two calibration methods, i.e. OCV \citep{Lam2022ax} and HATS \citep{Ooi2021b}. The calibrated stimuli were derived in a previous study \citep{Lam2022ax} with the exact hardware setup used in this study. 

\subsection{Objective assessment}

Taking reference from the minimum reporting requirements and analysis guidelines for soundscapes \citep{InternationalOrganizationforStandardization2018,InternationalOrganizationforStandardization2019}, suggested acoustic and psychoacoustic indicators were selected as a basis for objective comparison between calibration methods. Acoustic indicators include both A- and C-weighted equivalent continuous sound pressure levels \citep{InternationalOrganizationforStandardization2016d}, i.e. $L_\text{A,eq}$, $L_\text{C,eq}$. Psychoacoustic indicators include loudness \citep{InternationalOrganizationforStandardization2017a}, sharpness \citep{GermanInstituteforStandardisationDeutschesInstitutfurNormung2009b}, roughness \citep{EcmaInternational2020ECMA-418-2:2020PerceptionKenneth}, tonality based on Sottek's hearing model \citep{EcmaInternational2021ECMA-74EquipmentKenneth}, and fluctuation strength \citep{Fastl2001Kenneth}.

All the acoustic and psychoacoustic indicators were computed for each of the binaural channels across all 27 stimuli. In total, 3 sets of parameters were generated, namely of (1) the \textit{in-situ} binuaral recordings, (2) the binaural recordings of the in-situ tracks reproduced over headphones after calibration with the \textit{OCV} method, and (3) the binaural recordings of the in-situ tracks reproduced over headphones after calibration with the \textit{HATS} method. Each acoustic parameter was analysed as the mean across both binaural channels. Additionally, percentage exceedance levels (i.e. \{5, 10, 50, 90, 95\}\si{\%}) were also computed to quantify the variation of the indicators over time \citep{InternationalOrganizationforStandardization2016d}.

\subsection{Experimental design of subjective assessment}

To assess the perceptual differences between the calibration procedures, a within-subject experimental design was adopted. Participants would assess two sets of the same 27 stimuli, of which one was OCV-calibrated and the other other was HATS-calibrated. Each set was evaluated in separate sessions spaced at least one day apart to reduce the impact of memory effects. The participants were blind to the calibration method used in each session.

The stimuli were assessed based on the eight ISO/TS~12913-2:2018 perceived affective quality (PAQ) attributes, \textbf{in Bahasa Melayu} (ISO 639-3: \textsc{zsm}), on a 101-point scale from \textit{strongly disagree} (0) to \textit{strongly agree} (100). The Bahasa Melayu candidate translations used for the subjective assessment for the PAQ attributes were 
``menyenangkan'' for ``pleasant'', 
``huru-hara'' for ``chaotic'', 
``rancak'' for ``vibrant'', 
``uneventful'' for ``tidak meriah'',
``calm'' for ``tenang'', `
`annoying'' for ``membingitkan'', 
``eventful'' for ``meriah'', 
and ``monotonous'' for ``membosankan''
were respectively \citep{Aletta2020Kenneth,Lam2022bx}.

The sliders for all the PAQ attributes were set to a default neutral position (50) and were only activated after the stimuli had been played in their entirety at least once. Participants were allowed to repeat listening to each stimulus as many times as required. The experiment was conducted through a MATLAB-based graphical user interface in a quiet room \citep{Ooi_GUI_for_SATP}, wherein the audio stimuli were presented over studio monitor headphones (DT 990 Pro, Beyerdynamic GmbH \& Co. KG, Heilbronn, Germany) powered by an audio interface (UltraLite AVB, MOTU, Inc., Cambridge, MA, USA).  

Formal ethical approval was sought from the Institutional Review Board (IRB) of NTU (Ref. IRB-2021-293) for the responses collected. In compliance with ethical procedures, informed consent was obtained from all the participants, and also from the parent/guardian if the participant was between 18 and 20 years old. 

\subsection{Participants}

A total of 36 participants [female: 17 (\SI{47}{\%}); male: 19 (\SI{53}{\%})] were recruited for this study. However, 2 participants were excluded from the study. One participant was excluded due to failing a standard hearing test (mean threshold of hearing $< \SI{20}{\decibel}$~HL at $0.5, 1, 2, \SI{4}{\kilo\hertz}$) administered to all participants via an audiometer (AD629, Interacoustics A/S, Middelfart, Denmark), and another participant was excluded due to technical issues with the setup occurring within the duration of the experiment that could have led to results being unreliable. Hence, responses from a total of 34 participants [female: 16 (\SI{47}{\%}); male: 18 (\SI{53}{\%})] were used for the final data analysis. The ages of these 34 participants ranged from 19 to 45 years ($\Bar{x}_\text{age}=25.7$ years, $\text{SD}_\text{age}=6.7$, $\text{SE}_\text{age}=1.17$).

\begin{figure}[t]
    \centering
    \includegraphics[width=0.4\textwidth]{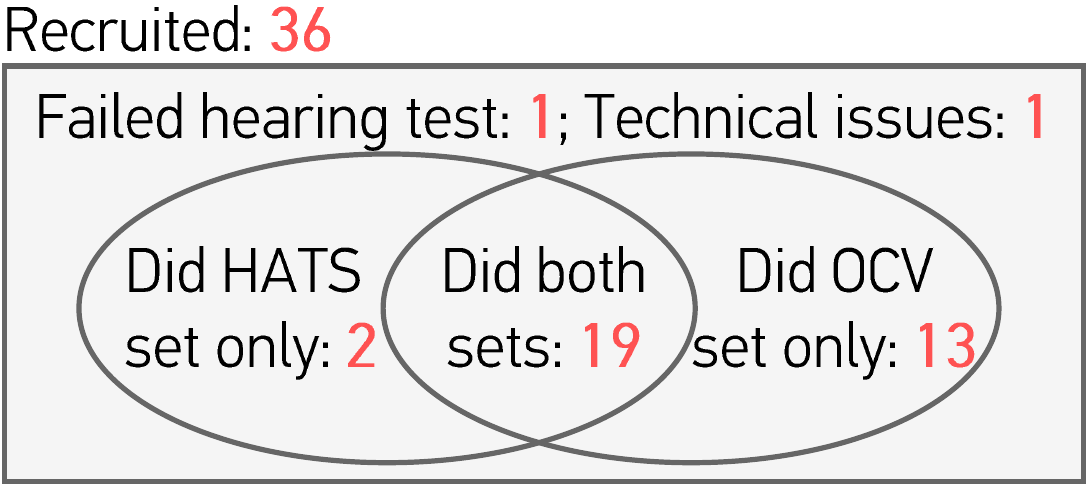}
    \caption{Participant breakdown for subjective assessment of OCV and HATS calibration methods.}
    \label{fig:participant_breakdown}
\end{figure}

In line with ethical guidelines, participants were allowed to terminate the study at any time, and thus only 19 participants [female: 10 (\SI{53}{\%}); male: 9 (\SI{47}{\%})] took part in both phases. 13 [female: 5 (\SI{38}{\%}); male: 8 (\SI{62}{\%})] participants completed the OCV set only, and 2 participants [female: 1 (\SI{50}{\%}); male: 1 (\SI{50}{\%})] completed the HATS set only. A graphical representation of the participant breakdown is shown in \Cref{fig:participant_breakdown}.

\subsection{Data analysis} \label{sec:dataAnalysis}

To account for the unpaired samples when performing the paired comparison between the OCV and HATS calibration methods across each stimulus and PAQ attribute pair, an optimal pooled $t$-test (OPTT) was adopted~\cite{Guo2017,Henriksen2018}.

The acoustic and psychoacoustic indicators were computed with a commercial software package (ArtemiS \textsc{suite}, HEAD acoustics GmbH, Herzogenrath, Germany). Bland-Altman (BA) parametric statistics and plots were generated to examine the agreement between in situ, and OCV and HATS calibrations methods \citep{MartinBland1986}. Additionally, \SI{95}{\%} confidence intervals were computed for the mean to examine the level of bias, and also for both the upper and lower bounds of the \SI{95}{\%} limits of agreement (LoA) to discover outliers.

All data analyses were conducted with the R programming language \citep{RCoreTeam2021} on a 64-bit ARM environment. The data that support the findings of this study are openly available in NTU research data repository DR-NTU (Data) at \href{https://doi.org/10.21979/N9/AUE2LL}{doi:10.21979/N9/AUE2LL}, and the replication code is available on GitHub at \href{https://github.com/ntudsp/ica22-calibration}{github.com/ntudsp/ica22-calibration}.

\section{Results and Discussion}

\subsection{Objective agreement between the OCV and HATS calibration methods}\label{sec:agreement}

Any bias introduced by the calibration methods during reproduction can be observed through the mean of the differences with the in-situ track across all stimuli in the BA plots. To determine outliers, the standard error was adopted as the limits of agreement (LoA), i.e., $\Bar{d}+z_{0.025}\cdot s_{d}$ and $\Bar{d}+z_{0.975}\cdot s_{d}$ for the lower and upper limits respectively, where $z_{0.025}=-1.96$ and $z_{0.975}=1.96$. $\Bar{d}$ and $s_{d}$ are the mean and standard deviation of the differences, respectively.

The calibration method can be considered unbiased if the line of equality (i.e. zero difference) is within the \SI{95}{\%} confidence intervals of the mean. Similarly, outliers are determined by differences that fall outside the lower and upper \SI{95}{\%} confidence intervals of the respective lower and upper LoAs. The two sided \SI{95}{\%} confidence intervals are determined by the $t$ standard error at $K-1$ degrees of freedom for small $K$ as
\begin{equation} \label{eq:tse}
   \pm t_{0.975}\cdot
   \left(\frac{1}{K}+\frac{1.96^2}{2\left(N+1\right)}\right)^{1/2}\cdot
   \left(\frac{s_{d}}{\sqrt{K}}\right),
\end{equation}
where $t_{0.975}=2.056$, and $K$ is the sample size \citep{Bland1999}.


\begin{figure}[t]
    \centering
    \includegraphics[width=0.95\textwidth]{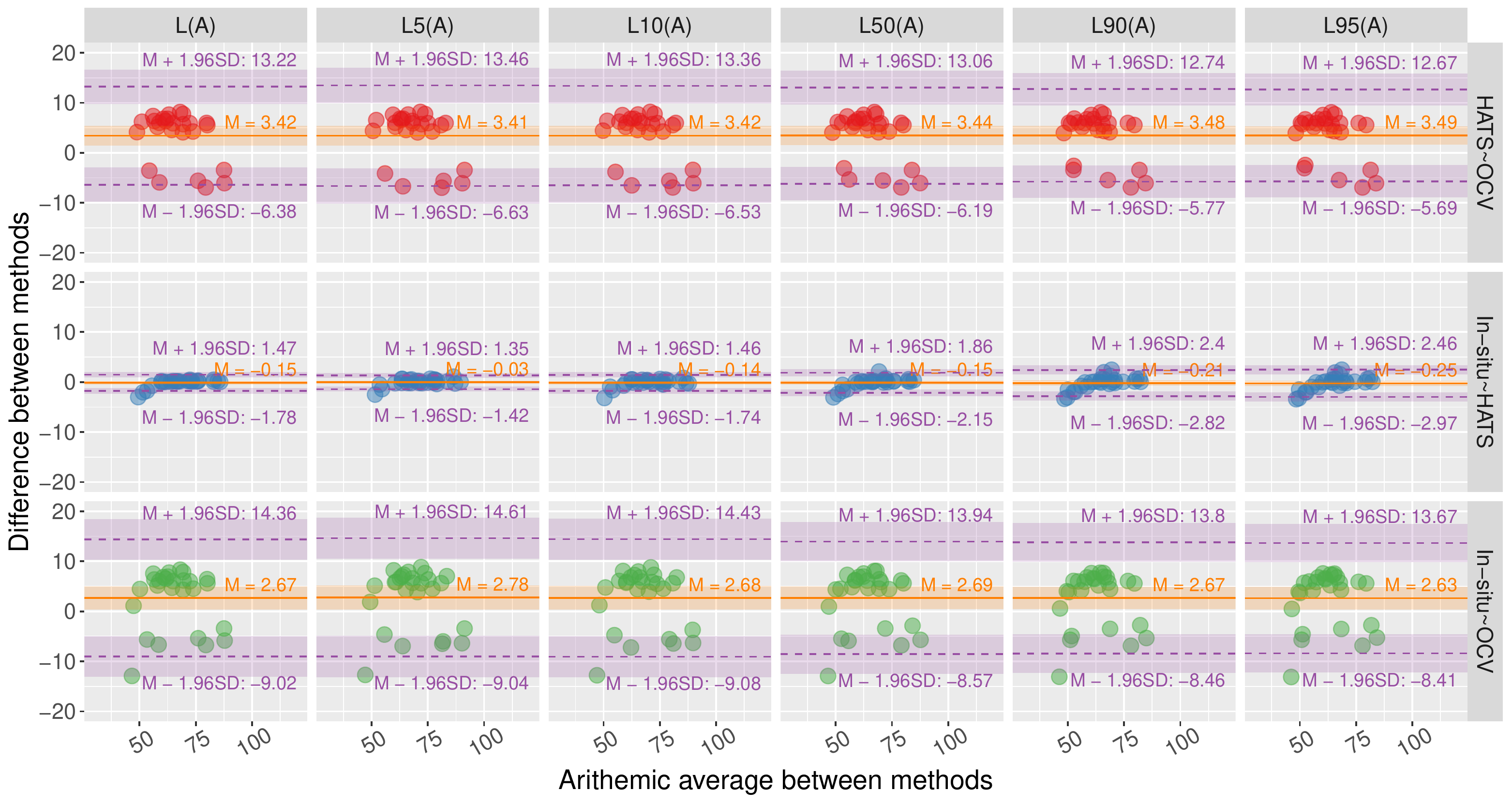}
    \caption{Bland-Altman plot of agreement between calibration methods across A-weighted level parameters in \si{\decibelA}. The average bias is estimated by the mean of the differences and bounded by the \SI{95}{\%} limits of agreement (LoA). Confidence intervals are based on \SI{95}{\%} confidence level for both LoAs (\textcolor{loaclr}{$\blacksquare$}) and the mean (\textcolor{mclr}{$\blacksquare$}). The numbers 5, 10, 50, 90, and 95 in the column headers indicate the exceedance levels.}
    \label{fig:BA_LA}
\end{figure}

Since the HATS calibration method algorithmically adjusts the $L_\text{A,eq}$ levels of the stimuli to within \SI{0.5}{\decibel} tolerance of the in situ $L_\text{A,eq}$ levels \citep{Ooi2021b}, it is expected that the differences between HATS and in situ across all tracks are near zero, as shown in \Cref{fig:BA_LA}. The narrow intervals were also observed across all A-weighted percentage exceedance levels. Moreover, no bias and outliers were present. However, the OCV method produced biases of about 
$\Bar{d}_{(\text{HATS}\sim\text{OCV})}\approx\SI{3.5}{\decibelA}$ and 
$\Bar{d}_{(\text{in-situ}\sim\text{OCV})}\approx\SI{2.7}{\decibelA}$ 
across all A-weighted SPL indicators in the respective differences between both HATS and in situ with OCV. The stimulus \texttt{KT01} was an outlier across all A-weighted SPL indicators in differences between in situ and OCV, with $d^{\text{KT01},\ L_\text{A,eq}}
_{(\text{in-situ}\sim\text{OCV})}
=\SI{-12.9}{\decibelA}$. It is worth noting that due to the high noise floor of the HATS signal chain, the track \texttt{KT01} failed to calibrate and was omitted from all the BA plots showing a comparison with the HATS calibration method.


Unsurprisingly, results were similar across the C-weighted SPL ($L_\text{C,eq}$) indicators, with only a slight bias of about $\Bar{d}_{(\text{in-situ}\sim\text{HATS})}\approx\SI{1.2}{\decibelC}$ in the differences between in situ and HATS, and biases of about $\Bar{d}_{(\text{HATS}\sim\text{OCV})}\approx\SI{2.5}{\decibelC}$ and $\Bar{d}_{(\text{in-situ}\sim\text{OCV})}\approx\SI{3}{\decibelC}$ in the respective differences between both HATS and in situ with OCV, and a clear outlier in \texttt{KT01} with $d_{(\text{in-situ}\sim\text{OCV})}^{\text{KT01},\ L_\text{C,eq}}=\SI{-13.5}{\decibelC}$. 


\begin{figure}[tb]
    \centering
    \includegraphics[width=0.95\textwidth]{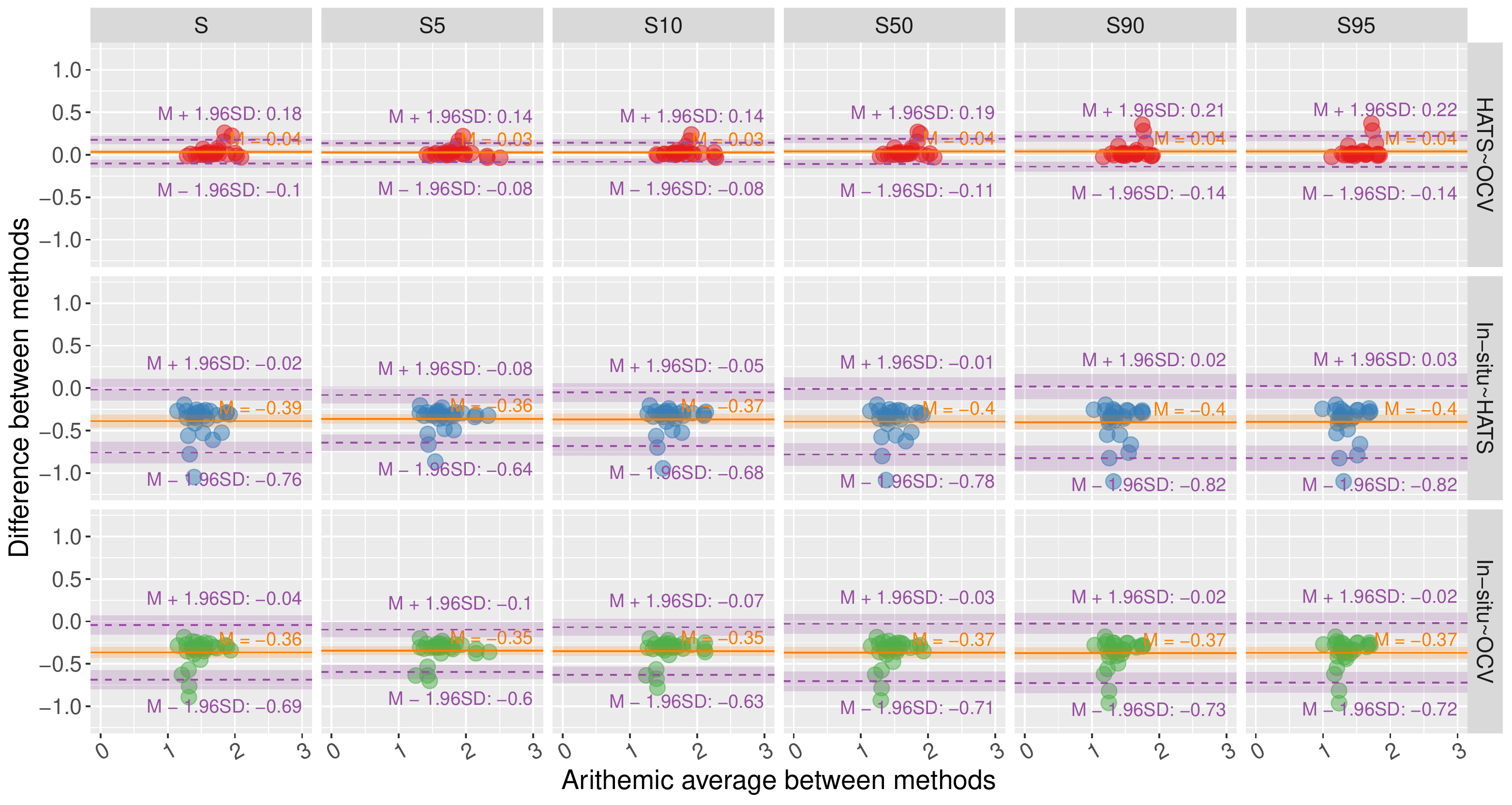}
    \caption{Bland-Altman plot of agreement between calibration methods across psychoacoustic sharpness parameters in acum. The average bias is estimated by the mean of the differences and bounded by the \SI{95}{\%} limits of agreement (LoA). Confidence intervals are based on \SI{95}{\%} confidence level for both LoAs (\textcolor{loaclr}{$\blacksquare$}) and the mean (\textcolor{mclr}{$\blacksquare$}). The numbers 5, 10, 50, 90, and 95 in the column headers indicate the exceedance levels.}
    \label{fig:BA_S}
\end{figure}

\begin{figure}[tb]
    \centering
    \includegraphics[width=0.95\textwidth]{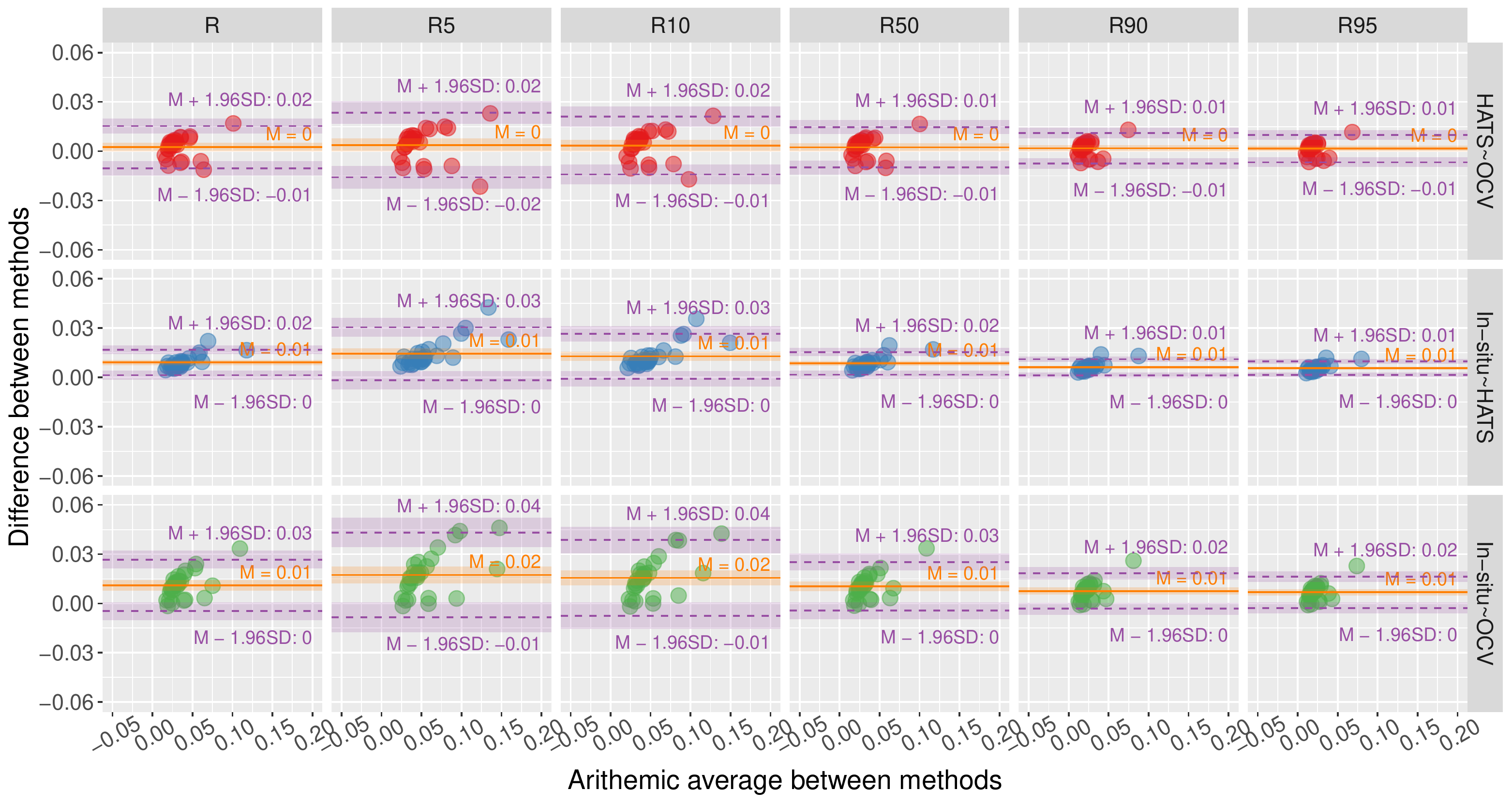}
    \caption{Bland-Altman plot of agreement between calibration methods across psychoacoustic roughness parameters in asper. The average bias is estimated by the mean of the differences and bounded by the \SI{95}{\%} limits of agreement (LoA). Confidence intervals are based on \SI{95}{\%} confidence level for both LoAs (\textcolor{loaclr}{$\blacksquare$}) and the mean (\textcolor{mclr}{$\blacksquare$}). The numbers 5, 10, 50, 90, and 95 in the column headers indicate the exceedance levels.}
    \label{fig:BA_R}
\end{figure}

Psychoacoustic loudness ($N$) indicators also exhibited similar trends to the weighted SPL indicators, whereby biases of about $\Bar{d}_{(\text{HATS}\sim\text{OCV})}\approx\SI{3}{\soneGF}$ and $\Bar{d}_{(\text{in-situ}\sim\text{OCV})}\approx\SI{2}{\soneGF}$ were observed in both the HATS and in situ differences with OCV, respectively. The stimulus \texttt{CT308} was the clear outlier in both the HATS ($\Bar{d}_{(\text{HATS}\sim\text{OCV})}^{\text{CT308},\ N}=\SI{-28.6}{\soneGF}$) and in situ ($d_{(\text{in-situ}\sim\text{OCV})}^{\text{CT308},\ N}=\SI{-27.5}{soneGF}$) differences with OCV. Owing to the similarities with the A-weighted SPL indicators, both the BA plots of the C-weighted SPL indicators and psychoacoustic loudness indicators were omitted for brevity.


The BA plots of the psychoacoustic sharpness indicators ($S$) indicated similar biases introduced by both the HATS and OCV calibrations, as shown in \Cref{fig:BA_S}. A clear outlier in the stimulus \texttt{VP01b} was observed in both HATS ($d_{(\text{in-situ}\sim\text{HATS})}^{\text{VP01b},\ S}=\SI{-1.05}{\acum}$) and OCV ($d_{(\text{in-situ}\sim\text{OCV})}^{\text{VP01b},\ S}=\SI{-0.893}{\acum}$) differences with in situ. The differences were similar across all percentage exceedance levels of sharpness. The absence of bias and narrow intervals between HATS and OCV suggests that the effect of sharpness could be attributed to the acoustic effects of the HATS signal chain. Based on inspection of the frequency spectrum, there was an overall increase in the noise floor, which could be attributed to the inability of the headphone in reproducing the dominant low-frequency content in \texttt{VP01b}.  


Biases in psychoacoustic roughness $R$ were also introduced by both HATS and OCV calibration methods, as shown in \Cref{fig:BA_R}. The bias of $\Bar{d}_{(\text{in-situ}\sim\text{HATS})}=\SI{0.01}{\asper}$ was consistent throughout all roughness indicators in the differences between in situ and HATS, whereas the biases observed in the \SI{5}{\%} and \SI{10}{\%} exceedance levels were double (i.e. $\Bar{d}_{(\text{in-situ}\sim\text{HATS})}=\SI{0.02}{\asper}$) of that in the other roughness indicators in the differences between in situ and OCV. Moreover, \texttt{CT308} ($d_{(\text{in-situ}\sim\text{HATS})}^{\text{CT308},\ R}=\SI{0.022}{\asper}$) was found to be the outlier in differences between in situ and HATS, whereas \texttt{W09} ($d_{(\text{in-situ}\sim\text{OCV})}^{\text{W09},\ R}=\SI{0.0334}{\asper}$) was the outlier in the difference between in situ and OCV across all roughness indicators. No biases nor outliers were observed in the differences between HATS and OCV in all roughness parameters. 


Since no bias and outliers were observed across both HATS and OCV methods in tonality and fluctuation strength indicators, the BA plots were omitted for conciseness.  All in all, any possible perceptual differences between the OCV and HATS calibration methods could be attributed to differences in weighted SPL indicators and psychoacoustic loudness. 

\subsection{Effect of calibration on perceived affective quality}

\begin{figure}[t]
    \centering
    \includegraphics[width=0.95\textwidth]{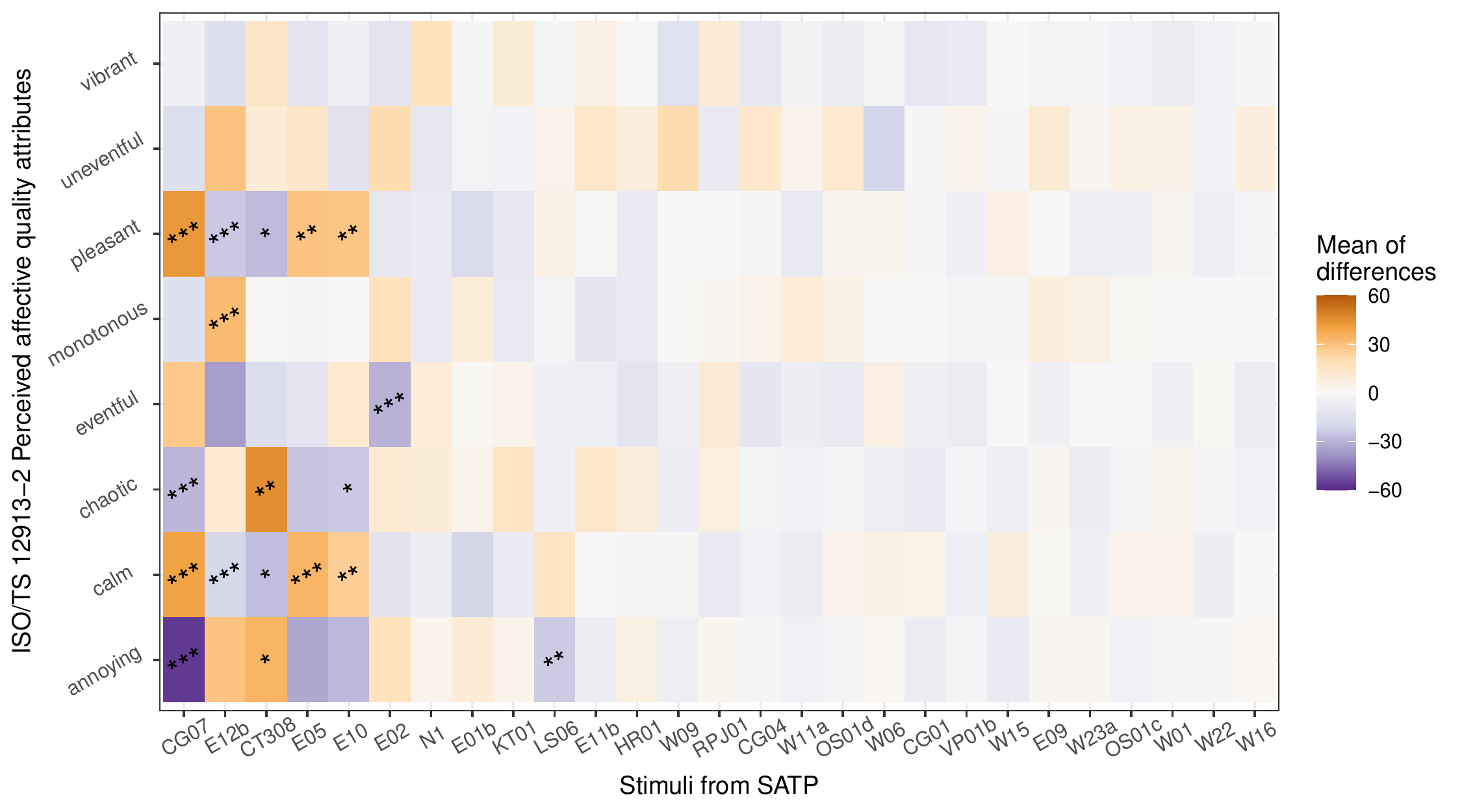}
    \caption{Heatmap indicating the estimated mean of the differences between the OCV and HATS calibration for each stimulus-attribute pair based on the optimal pooled t-test with Bonferroni correction. Significance levels at \{$5, 1, 0.1,0.01$\}\si{\%} are symbolized by \{$\ast$, $\ast\ast$, $\ast\ast\ast$, $\ast\ast\ast\ast$\}, respectively. Stimuli are ranked by the absolute mean differences across all attributes.}
    \label{fig:optt}
\end{figure}

A within-subject design was employed to investigate the perceptual differences between the HATS and OCV calibration methods. As mentioned in \Cref{sec:dataAnalysis}, presence of unpaired samples due to participant dropout was managed by adopting a \textit{t}-test for partially paired data, i.e. optimal pooled \textit{t}-test (OPTT) \cite{Guo2017}. A total of $N_{stimuli}\times N_{PAQ}=27\times 8=216$ stimuli-attribute paired comparisons were made, and with Bonferroni corrections applied to the resultant \textit{p}-values. The mean of the differences between the calibration methods for each pair, as well as the significance of the adjusted \textit{p}-values are visualized in a heatmap, as shown in \Cref{fig:optt}. 

\textit{Calmness} and \textit{pleasantness} were significantly different in 5 stimuli (i.e. \texttt{CG07}, \texttt{E12b}, \texttt{CT308}, \texttt{E05}, \texttt{E10}). Except for  \texttt{E05}, four of those stimuli had significant differences in at least one other attribute. For \texttt{LS06} and \texttt{E02}, significant differences were only observed in \textit{annoying} and \textit{eventful}, respectively. The differences in weighted SPL and psychoacoustic loudness indicators did not seem to affect all the stimuli equally. For instance, stimuli with the largest differences in A-weighted SPL (e.g. \texttt{W23a}, \texttt{CG01}, \texttt{W11a}, \texttt{N1}), C-weighted SPL (e.g. \texttt{CG01}, \texttt{W11a}, \texttt{OS01d}), and loudness (e.g. \texttt{W09}, \texttt{OS01d}, \texttt{E11b}) were not amongst those stimuli that registered significant perceptual differences. Hence, interaction effects between indicators could be present and warrants further investigation. 

\section{Conclusion}

A proposed low-cost OCV calibration procedure for reproduction of audio stimuli over headphones was examined objectively and perceptually, in comparison with a calibrated HATS, in the context of ISO 12913 soundscape evaluations. The OCV method differed from the \si{\decibelA}-based HATS method in objective weighted SPL, loudness, $R5$ and $R10$ indicators, but was similar in sharpness, tonality, fluctuation strength, and the rest of the roughness indicators across 27 distinct audio stimuli. Significant perceptual differences were observed for a small number via paired comparison analysis of stimuli that did not immediately correspond to specific objective differences.

However, the interpretations of the perceptual effects could be modulated by the small sample size, unpaired samples, and inherent intricacies due to translation of PAQ attributes from English to Bahasa Melayu. Nevertheless, the significant differences observed in the main axis PAQ attributes provides preliminary evidence of the potential perceptual differences due to calibration methods for laboratory assessments of soundscape. 

\section*{Acknowledgements}
This research is supported by the Singapore Ministry of National Development and the National Research Foundation, Prime Minister's Office under the Cities of Tomorrow Research Programme (Award No. COT-V4-2020-1). Any opinions, findings and conclusions or recommendations expressed in this material are those of the authors and do not reflect the view of National Research Foundation, Singapore, and Ministry of National Development, Singapore.

\renewcommand{\bibsep}{0pt}
\renewcommand{\bibnumfmt}[1]{\makebox[0.5cm][l]{#1.}}
\bibliographystyle{vancouver}
\bibliography{references_bhan, references_kenneth}

\end{document}